\documentclass[runningheads]{llncs}
\usepackage{graphicx}
\begin{document}
\title{Coronary Artery Segmentation in Angiographic Videos Using A 3D--2D CE--Net}
\author{Lu Wang\inst{1} \and
Dong-xue Liang\inst{1,*} \and
Xiao-lei Yin\inst{1} \and
Jing Qiu\inst{1} \and
Zhi-yun Yang\inst{2} \and
Jun-hui Xing\inst{3} \and
Jian-zeng Dong\inst{2,3} \and
Zhao-yuan Ma\inst{1}}
\authorrunning{Lu Wang, Dong-xue Liang, et al.}
\institute{1 The Future Laboratory, Tsinghua University, Beijing 100084, China.\\
2 Center for Cardiology, Beijing Anzhen Hospital, Capital Medical University, Beijing 100029, China.\\
3 The First Affiliated Hospital of Zhengzhou University, Zhengzhou 450052, China.\\
** Dong-xue Liang is the correspondence author (e-mail: liang\_laurel@tsinghua.edu.cn).}
\maketitle              % typeset the header of the contribution
\begin{abstract}
Coronary angiography is an indispensable assistive technique for cardiac interventional surgery. Segmentation and extraction of blood vessels from coronary angiography videos are very essential prerequisites for physicians to locate, assess and diagnose the plaques and stenosis in blood vessels. This article proposes a new video segmentation framework that can extract the clearest and most comprehensive coronary angiography images from a video sequence, thereby helping physicians to better observe the condition of blood vessels. This framework combines a 3D convolutional layer to extract spatial--temporal information from a video sequence and a 2D CE--Net to accomplish the segmentation task of an image sequence. The input is a few continuous frames of angiographic video, and the output is a mask of segmentation result. From the results of segmentation and extraction, we can get good segmentation results despite the poor quality of coronary angiography video sequences.
\end{abstract}

\section{Introduction}
Physicians have been undertaking interventional surgeries to diagnose and treat cardiovascular diseases for quite a few years. They locate, assess and diagnose the blood vessel stenosis and plaques by directly watching the angiographic videos with naked eyes during the surgeries. Based on their experiences, the physicians quickly make a qualitative judgment on the patient's coronary artery condition and make the treatment plan. This direct method is greatly affected by human factors and lacks accuracy, objectivity and consistency. Automated cardiovascular segmentation will help reduce the diagnostic inaccuracies for physicians. Many blood vessel extraction methods based on image segmentation have emerged driven by this motivation. Recently, with the development of deep learning, various deep neural network architectures have been proposed and applied in the medical image processing tasks, especially in the image segmentation field \cite{ref_article1,ref_article2}. Early deep learning--based approaches use the image patches and a sliding window block to traverse the image \cite{ref_article3}. But the sliding window method casts a huge amount of computation, and misses the global contexts of the image at the same time. Since 2014, fully--convolutional network, encoder-decoder network and U--Net \cite{ref_article4} were proposed for medical image segmentation and achieved good results. After that many different methods based on U--Net architecture have sprung up. M--net added a multi--scale input image and deep supervision to the original U--Net architecture \cite{ref_article5}. New modules have been proposed to replace some blocks in the U--Net architecture to enhance the feature learning ability. Gibson et al., proposed a dense connection in the encoder block \cite{ref_article6}. To improve the segmentation performance, Zhao et al., introduce a modified U--Net by adding a spacial pyramid pooling \cite{ref_article7}. Gu et al., insert a dense atrous convolution block (DAC) and a residual multi--kernel pooling block (RMP) into the bottle-neck part of U--Net to extract and preserve more spatial context, and this it the state-of-art network design CE--Net \cite{Gu}.

Nevertheless, due to the particularity of cardiovascular angiographic images, some essential problems are encountered with in the cardiovascular segmentation tasks. First, the shape of the blood vessels in the angiographic images is complex and easily deformed. Blood vessels have a tubular curved structure, and some blood vessels can block, cover or entangle with one another, making the semantic information confusing in the images. Second, the angiographic images contain not only blood vessels, but also other organs and tissues. To make things worse, the shape and grayscale values of some tissues are similar to that of blood vessels, which makes it even more difficult to correctly extract the object. Third, due to the need for X--rays during angiographic imaging, the health of the patients and physicians is damaged. In order to minimize the damage, it is inevitable to reduce the dose of X--rays, which causes low illumination. As a consequence, it reduces the signal--to--noise ratio of the images, and makes the segmentation tasks even more challenging. Considering that the coronary artery angiographic video is a series of time--continuous image sequences, rather than individual images, combining and processing several consecutive frames of images may provide a good idea and insight for solving these problems. For instance, the blood vessels that are blocked by each other in one image may be separated in another image. The problem of low signal--to--noise ratio caused by low illumination may be eliminated by the accumulation of multiple images. Therefore, the temporal dimension of the video is also important. Using spatial and temporal information to segment blood vessels in angiographic videos becomes a subject worthy of study.

In the mean time, since 2D CNNs has achieved good results in image classification and segmentation, researchers have reoriented their interests to the video field. In the classification and segmentation tasks of videos, researchers used temporal information as the third dimension and introduced the concept of 3D convolutional neural networks. The Inception 3D (I3D) architecture proposed by Carreira et al., is one of the earliest 3D CNN models \cite{ref_article8}. This model inflates all 2D convolutional kernels in the Inception V1 architecture \cite{ref_article9} into 3D kernels, and is trained on the large--scale Kinetics dataset of videos\cite{ref_article10}. However, the computational cost of 3D convolution is extremely high, so a variety of mixed convolution models based on ResNet architecture have been proposed to resolve this dilemma. Another way to reduce the computational cost is to replace the 3D convolution with separable convolutions. In order to effectively make use of the temporal dimension, \cite{ref_article11} proposed a R(2+1)D convolutional neural network, and \cite{ref_article12} proposed a model called the pseudo 3D network. 

In this article, we consider combining the advantages of 3D and 2D convolution to accomplish the task of blood vessel segmentation from coronary artery angiographic videos. Based on the architecture of 2D U--Net and its derivative CE--Net, we propose a 3D--2D network. The 3D convolutional layer serves to extract and process temporal information in the video, and the 2D CE--Net extracts the spatial information. Our main contributions are as follows:
\begin{itemize}
	\item A novel deep learning architecture combining 3D and 2D convolution to extract both temporal and spatial information from videos.
	\item A new application of deep learning--based video segmentation algorithm in the medical imaging field.
\end{itemize}

\section{Methods}
\label{sec:methods}
\subsection{\label{sec:level2}3D convolution}

In this paper, 3D convolution is used to extract and fuse the time--domain context information of coronary artery angiographic video, combined with 2D CE--Net, to achieve blood vessel segmentation for each frame in the angiographic video. 3D convolution has been widely used in the field of volumetric medical image processing, such as tumor segmentation tasks of layered scanned organic CT images. However, CT scan imaging is fundamentally different from coronary angiography. What CT gets is a series of layered two-dimensional images. Layers are stacked into a three--dimensional volumetric image. The physical meaning of the spatial dimensions (width, height, depth) of this volumetric image is very clear, and the correlation between layers constitutes the spatial context information. Therefore, it is a natural processing method to apply 3D convolutional network to volumetric CT image segmentation. Coronary angiography obtains a two--dimensional image whose spatial depth information has been squeezed. There are only two spatial dimensions, ie., width and height. We cannot directly use 3D convolutional networks to process such images. We observe that coronary angiography is a continuous imaging process, and the sequence of moving contrast images forms a video. The traditional coronary artery segmentation task is to process independent contrast image frames, and a 2D convolutional neural network was used to extract the spatial details and context information inside the image. Considering that the 3D convolutional network can extract spatial and temporal context features of the video at the same time, we stack each frame of the coronary angiography imaging video to form a three-dimensional data, except that the depth dimension of the space is replaced by the time dimension. 3D convolutional networks have been widely used in the field of video processing to extract the spatial--temporal features and contextual information of videos. Previous work proved that the temporal context information in the videos contains rich image semantic representations. Making full use of the temporal information can greatly benefit image semantic segmentation, and enhance the performance of tasks such as scene understanding and target recognition.

\subsection{\label{sec:level2}Network architecture}

\begin{figure}[h]
\centering
\includegraphics[width=\textwidth]{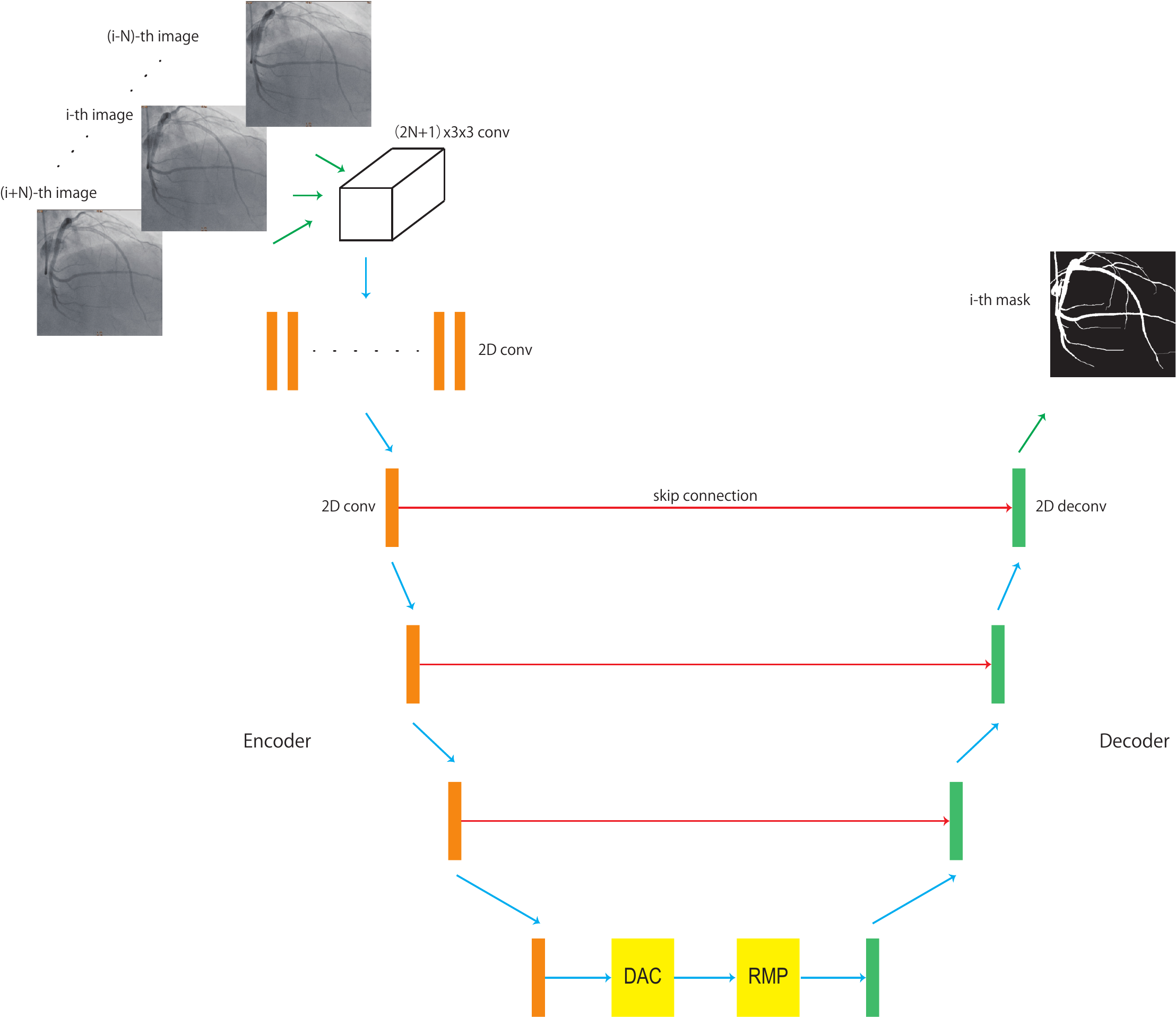}
\caption{Network architecture: the input layer of the network is a 3D convolutional layer which has a kernel size of \((2N+1)\times 3\times 3\), and it accepts \(2N+1\) adjacent frames and outputs 2D feature maps of \(16\) channels, which are further fused by a 2D convolutional layer. The subsequent part of the network is analogous to 2D CE--Net. The network outputs a single mask image corresponding to the central frame of the input. In practice, \(N\) is usually set to \(1\) or \(2\).}
\label{fig:network}
\end{figure}

This paper is inspired by the work of deep learning based video processing and 2D medical image segmentation methods. The input layer of CE--Net is replaced with a 3D convolution layer, so that the network accepts several frames of time continuous images as input. We define the image to be segmented as the target image, and the temporally adjacent \(2N\) images are referred to as auxiliary images, which contain \(N\) images before and after the target image, respectively. The \(2N+1\) input images are extracted and fused by the 3D convolution layer to obtain several channels of fused 2D feature maps, with the same resolution of the input images. Another 2D convolution layer is inserted after the input 3D layer, to adjust the channel number of the feature maps to fit the requirement of the CE--Net. Empirically, the number of the output channels of the 3D convolution layer is 16, and the subsequent 2D convolution layer accepts 16 channels as input and outputs 2D feature maps of 3 channels. After processes through the encoder, bottle neck, decoder, and skip connections of the subsequent CE--Net, a two-dimensional image is eventually output as the segmentation result of the target image. As the heart beats cyclically, the cardiovascular vessels expand and contract in cycles accordingly. In the process of vessel motion, mutual blocking between blood vessels and deformations will inevitably occur. The value of \(N\) is an important hyperparameter. The smaller \(N\) is, the fewer input neighboring images are captured, and the extracted temporal context information is confined to a small range, but more accurate spatial details are retained. The larger the value of \(N\) is, the more adjacent images are input into the network, and the broader the temporal context information extraction and fusion is. It keeps a long--term trend and has a positive significance for eliminating the occlusion of vessels due to vascular motion. Considering that the imaging frame rate of existing coronary angiography equipments is relatively ordinary, generally around 15 frames per second, and the cardiac cycle is generally smaller than one second, we set the value of \(N\) as 1 or 2, that is, the network accepts 3 or 5 adjacent frames as input. The network architecture and its input and output are illustrated in Fig.~\ref{fig:network}.

\section{\label{sec:level1}Experiments}

\subsection{Dataset}
In collaboration with several hospitals, we collected \(170\) coronary artery angiography video clips of cardiac interventional surgeries. The length of each video clip ranges from \(3\) to \(10\) seconds, recording the process from the injection of the contrast agent to the flow, diffusion and gradual disappearance of the contrast agent in the blood vessels. These videos are desensitized to hide the patient's personal information and protect the patient's right to privacy. The acquisition equipments of these video clips are products of many different manufacturers, so the image resolution ranges from \(512\times512\) to \(769\times718\), the frame rate ranges from \(10fps\) to \(15fps\), and the original video format is wmv, MP4, etc. We use ffmpeg software toolkit to extract each frame of the videos and save it as a losslessly encoded RGB image file with a color depth of \(8\) bits in each chromatic channel. The total number of the source images is 8,835. We invite students from several medical colleges to manually segment and label the coronary arteries in these source images to generate a label image with an identical resolution of each image. Thus we have obtained a dataset containing 8,835 source images and 8,835 label images, which is called the SetV. 

\subsection{Partition of the training and test sets}
The SetV dataset is comprised of 170 sub--paths, each containing an average of 52 source images and 52 label images, and the images are stored in the chronological order of each video clip. We divide the images of each sub-path into a training set and a test set, where the training set accounts for \(5/6\) of the total images and the test set accounts for the other \(1/6\). As the input layer of the network accepts several time--continuous images, the division method is no longer a random selection of a single image, but the first \(1/6\) of each subpath is used as the test set, and the last \(5/6\) is used as the training set, or in the reverse order. The two division methods are randomly determined when the training process begins. This hard--cutting partitioning method avoids the error that an image appears in both the training and test sets. In order to ensure that each image is used in training or test, we have padded the training and test sets according to the value of \(N\) in the network input layer. The first image and the last image in each subset are copied \(N\) times to achieve a similar effect as the padding in the convolutional operations.

\subsection{Performance metrics}
In training the model, we use the combined loss of dice loss and regularization loss as the objective function. We also calculate the IOU (intersection over union) to evaluate the performance in both training and test stages. Dice coefficient and IOU are both metrics on the similarity between two sets, with slight differences,
\begin{eqnarray}
dice\_coe = \frac{2|X\bigcap Y|}{|X|+|Y|},\\
L\_{dice} = 1 - dice\_coe,\\
L = L\_{dice} + L\_{reg},\\
IOU = \frac{|X\bigcap Y|}{|X\bigcup Y|},
\end{eqnarray}
where \(|X|\) denotes calculation of the number of elements in a set \(X\). We also calculate the sensitivity (true positive rate, TPR), and the specificity (true negative rate, TNR) of the model on the test set.

\subsection{Experimental results}
\begin{figure}[h]
\centering
\includegraphics[width=0.75\textwidth]{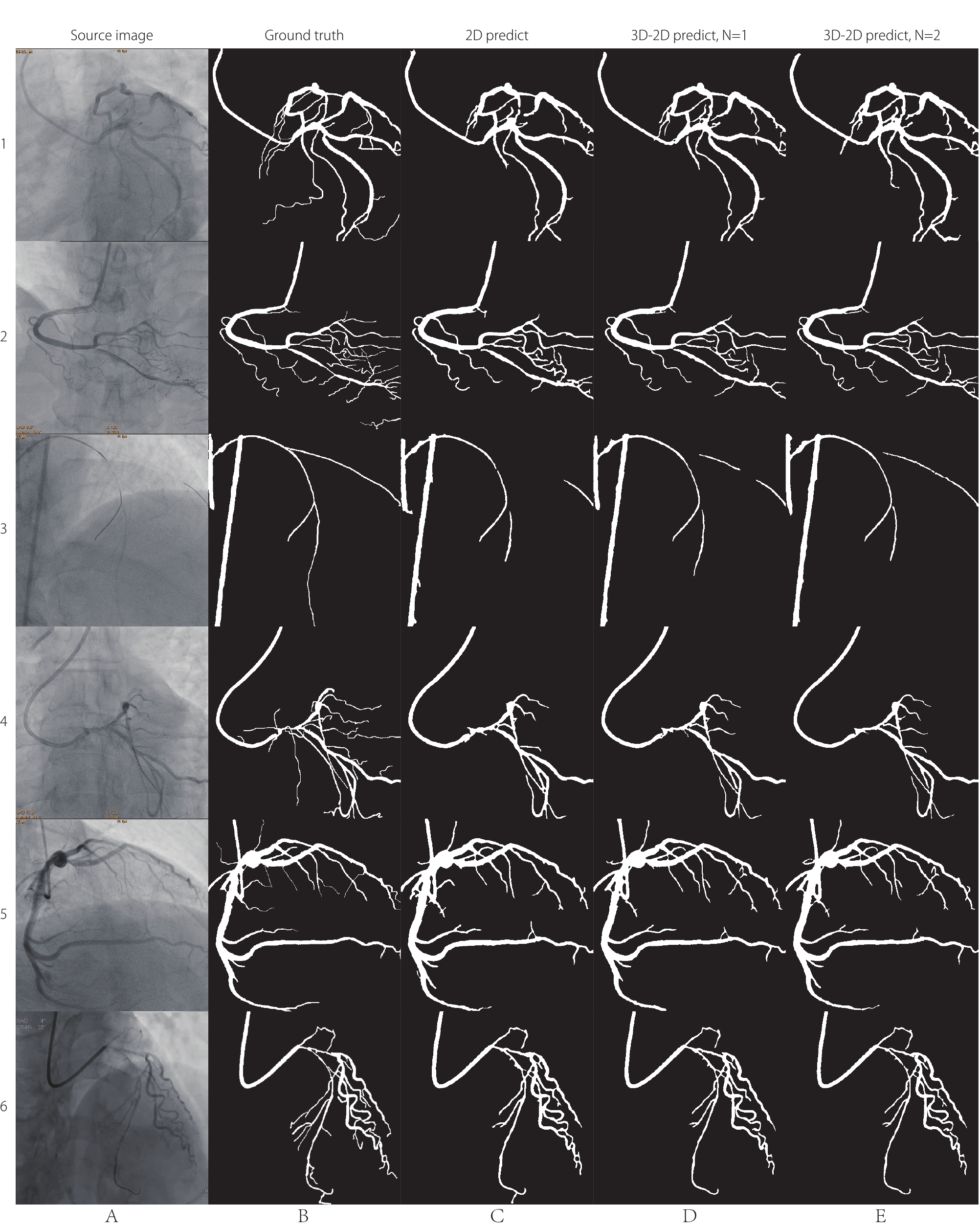}
\caption{Six examples of blood vessel segmentation from coronary artery angiographic videos/images. Row 1\(\sim\)6 correspond to six video clips, respectively. Column A: a target image captured from the video; col. B: ground truth, manually labeled blood vessels; col. C: segmentation result from 2D CE--Net; col. D: segmentation result from our 3D--2D CE--Net, where \(N\) is 1; col. E: segmentation result from our 3D--2D CE--Net, where \(N\) is 2.} \label{fig:result}
\end{figure}

All the source and label images are resized to the resolution of \(448 \times 448\) before being fed into the network. The four encoder layers of the 2D CE--Net are initiated with a Resnet--34 model pre--trained on the public ImageNet dataset. Several modifications are made to the CE--Net structure, for instance, we replace the batch normalization by instance normalization in both the encoder and decoder parts, and use stochastic gradient descent as the optimization method. The learning rate is fixed to \(2\times 10^{-4}\), and the batch size is 4. We set the parameter \(N\) to 1 or 2, ie., the number of input neighbor frames to 3 or 5, and train the model in 100 epochs. The output mask images are post-processed to remove the small connected components in each image and save the major parts of the mask as the segmentation result. As a comparison, we also test the original 2D CE--Net on our dataset. Some results are illustrated in Fig. \ref{fig:result}.

Compared with the segmentation results from the state-of-art technique, ie., the 2D CE--Net (column C in Fig.~\ref{fig:result}), our 3D--2D models (column D and column E)  demonstrate better preservation capability of both the vascular details and global structures. The performances of the three models are listed in Table~\ref{tab1}. Our 3D--2D models outperform the state-of-art technique in all the three metrics. With the increasing number of the time--continuous images input into the 3D convolutional layer, the performance of the segmentation improves in a noticeable degree.

\begin{table}
\centering
\caption{Comparison of the performances of 2D CE--Net, 3D--2D CE--Net with \(N=1\) and \(N=2\) on the test set of SetV.}\label{tab1}
\begin{tabular}{|l|l|l|l|l|l|}
\hline
Model &  \(N\) & IOU & sensitivity & specificity \\
\hline
2D CE--Net & & 0.795 & 0.762 & 0.9931 \\
\hline
3D--2D CE--Net & 1 & 0.813 & 0.774 & 0.9939 \\
\hline
3D--2D CE--Net & 2 & 0.818 & 0.779 & 0.9941 \\
\hline
\end{tabular}
\end{table}

\section{Conclusion}
By adding a 3D convolutional layer to the input layer of the 2D CE--Net to extract and fuse the temporal information in the coronary artery angiographic video, we obtain a better performance in the segmentation tasks of the blood vessels. The experiments demonstrates that the more adjacent frames are fed into the network, the better the performance is. This work shows that the time--domain information of videos has practical significance for image segmentation and interpretation, and is worthy of further study.

\bibliographystyle{splncs04}

\end{document}